\documentclass[superscriptaddress, prl, twocolumn]{revtex4}

\usepackage{times,epsfig,wrapfig}
\usepackage{amsmath}
\usepackage{amssymb}

\renewcommand\Re{\operatorname{Re}}
\renewcommand\Im{\operatorname{Im}}

\begin{document}

%%% Title of paper

\title{Surface plasmon-polariton resonance at diffraction of THz radiation on semiconductor gratings}

\author{I. S. Spevak}
\email{stigan@rambler.ru}
\affiliation{O. Ya. Usikov Institute for Radiophysics and Electronics of the National Academy of Sciences of Ukraine., Ac. Proskury 12, 61085 Kharkiv, Ukraine}
\author{M. Tymchenko}
\affiliation{O. Ya. Usikov Institute for Radiophysics and Electronics of the National Academy of Sciences of Ukraine., Ac. Proskury 12, 61085 Kharkiv, Ukraine}
\author{V. K. Gavrikov}
\affiliation{Institute of Radio Astronomy of the National Academy of Sciences of Ukraine, Chervonopraporna 4, 61002 Kharkiv, Ukraine}
\author{V. M. Shulga}
\affiliation{Institute of Radio Astronomy of the National Academy of Sciences of Ukraine, Chervonopraporna 4, 61002 Kharkiv, Ukraine}
\author{J. Feng}
\affiliation{Jilin University, 2699 Qianjin Str., Changchun 130012, China}
\author{H. B. Sun}
\affiliation{Jilin University, 2699 Qianjin Str., Changchun 130012, China}
\author{Yu. E. Kamenev}
\affiliation{O. Ya. Usikov Institute for Radiophysics and Electronics of the National Academy of Sciences of Ukraine., Ac. Proskury 12, 61085 Kharkiv, Ukraine}
\author{A. V. Kats}
\affiliation{O. Ya. Usikov Institute for Radiophysics and Electronics of the National Academy of Sciences of Ukraine., Ac. Proskury 12, 61085 Kharkiv, Ukraine}

\date\today{}

\begin{abstract}
    Resonance diffraction of THz HCN laser radiation on a semiconductor (InSb) grating
    is studied both experimentally and theoretically. The specular reflectivity suppression
    due to the resonance excitation of the THz surface plasmon-polariton is observed
    on a pure semiconductor grating and on semiconductor gratings covered with a thin striped layer
    of the residual photoresist. Presence of a thin dielectric layer on the grating surface
    leads to the shift and widening of the plasmon-polariton resonance.
    A simple analytical theory of the resonance diffraction on a shallow grating
    covered with a dielectric layer is presented. Its results are in a good accordance with the experimental data.
    Analytical expressions for the resonance shift and broadening can be useful for sensing data interpretation.
\end{abstract}

\maketitle

The terahertz band ($0.3-10$ THz) is a very promising frequency range of the electromagnetic
spectrum due to a wide variety of possible applications such as imaging, nondestructive evaluation,
biomedical analysis, chemical characterization, remote sensing, (including detection of agents associated with illegal
drugs or explosives), communications, etc \cite{Tonouchi_2007, Chan_2007, Zhang_2010}.
Therefore, in spite of the lack of cheap and compact THz sources and detectors,
fundamental and applied researches in the THz area are a problem of today.

Sensing and detection capabilities of the THz radiation can be substantially enhanced by employing the surface
plasmon-polariton (SPP) owing to its high field concentration near a metal-dielectric interface. SPPs are transverse magnetic
waves that propagate along the boundary between the conducting and dielectric media and are coupled to the collective
oscillations of free electrons in a conductor \cite{Raether_1988}. In the visible and IR ranges metals are
usually considered as standard SPP supported media. In contrast to this, in the THz region metals
behave as perfect conductors that results in a large SPP extent into the adjacent dielectric (weak localization) and a huge free path length.
Actually, the terahertz SPP at a flat metal loses its surface nature and its existence is questionable.

Meanwhile, many semiconductors possess optical properties allowing efficient THz SPP excitation,
propagation and manipulation without any additional treatment \cite{Rivas_2004, Kuttge_2007, Balakhonova_2007, Spevak_2011}.
It is due to the fact that the dielectric permittivity of these semiconductors in THz is analogous to that of
metals in the visible spectrum range.
So, all effects observed in optics and associated with SPP, in particular,
total suppression of the specular reflection \cite{Hutley_1976, Gandelman_1983},
resonance polarization transformation \cite{Elston_1991_b, Kats_2002, Kats_2007},
enhanced transmission through optically thick structured metal layers
\cite{Ebbesen_1998, Garsia-Vidal_2010}, and its counterpart,
suppressed transmission through optically thin structured metal films
\cite{Spevak_2009, Braun_2009, Xiao_2010-B, D'Aguanno_2011},
can be realized in the THz region with application of SPP. An additional advantage of semiconductors
as plasmon-polariton supporting media is related to the fact that their optical parameters can be
controlled by optical excitation or thermal action
\cite{Rivas_2006_a, Sanches-Gill_2006, Rivas_2006_b}.

A majority of recent THz experiments are carried out using the common terahertz time-domain setup
\cite{Rivas_2004, Kuttge_2007, Fattinger_1989, Qu_2004, Gaborit_2009} with a pulse broadband THz source.
The typical full angle divergence of the THz beam is $\sim 0.1$ rad and the spectral resolution of the
setup is $\sim 6...18$ GHz  \cite{Nazarov_2007, Fattinger_1989, Qu_2004, Gaborit_2009}.
At the same time, the spectral half-width of the SPP resonance on a shallow
semiconductor gratings is usually about $\leq 5$ GHz,
and the angular half-width of the resonance is $\sim 0.01$ rad. Evidently, a more collimated THz beam
and better spectral resolution of the experimental setup are needed for detailed study of the SPP
resonance's  fine structure, and the THz laser is the most suitable tool for this purpose.

In this Letter we investigate the resonance diffraction of the HCN (hydrogen cyanide) laser
radiation (the wavelength $\lambda = 336.6$~$\mu$m) on the InSb grating (the dielectric permittivity \cite{Palik_1998}
$\varepsilon = -87 + 37.8i$). The laser beam quality is characterized
by the spectral width $< 10$ MHz; the full angular width is $\sim 0.01$ rad ($\sim 0.9$ deg).

\begin{figure}[!h]
    \scalebox{0.4}[0.4]{\includegraphics{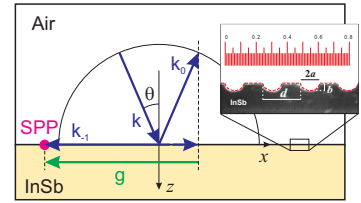}}
    \caption {Geometry of the diffraction problem and the grating profile.}
    \label{Geometry}
\end{figure}

We study four gratings produced successively by the
standard lithographic process from a single $0.8$~mm thick InSb wafer. At the beginning, the experiment
was carried out with the first grating without removing the photoresist striped pattern. Then, this grating
underwent subsequent etching resulting in
the second grating with deeper grooves, and so on. As a result, three successive gratings had the same period
$d = 254$~$\mu$m but different grooves' depth and width. The grooves had the semi-elliptic
profile with the semi-axes $(a,b)$ equal to $(53.5\mu\text{m}, 10\mu\text{m})$, $(65\mu\text{m}, 20\mu\text{m})$,
$(71.5\mu\text{m}, 24\mu\text{m})$, respectively (the width of the grooves equals $2a$, and their depth, $b$,
was counted from the semiconductor surface).
Finally, the fourth grating came out from the third one after removing the residual photoresist layer, i.e.
it had the same profile parameters, $(a,b) = (71.5\mu\text{m}, 24\mu\text{m})$ as the third grating.

The HCN laser radiation was $p$-polarized. The intensity distribution in the laser beam was nearly
Gaussian with the initial radius of $6.7$~mm.
The incidence plane was perpendicular to the grating's grooves and all diffraction orders also lied in
the incidence plane and were $p$-polarized, see  Fig.~\ref{Geometry}.

The diffraction geometry was chosen so that the resonance takes place in the $-1^{\text{st}}$ diffraction order.
In this case the only propagated diffraction order is the specular
reflected one. Owing to this, the radiative losses are minimal, resulting in maximizing the resonance strength and
minimizing the resonance width.

\begin{figure}[!t]
    \centerline{\scalebox{0.45}[0.45]{\includegraphics{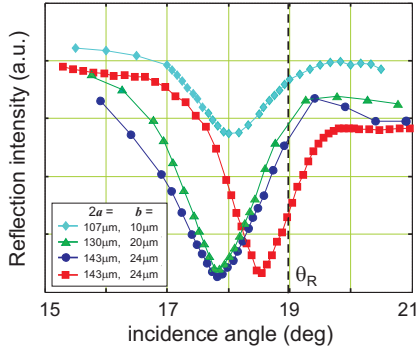}}}\caption[]
    {Experimental dependence of the specular reflectivity on the incidence angle.
    The grating parameters are indicated in the legend. The three initial curves
    correspond to the gratings with the residual photoresist. The fourth curve (red square markers)
    corresponds to the pure grating with the removed photoresist.}
    \label{Experiment_InSb}
\end{figure}

The experimental results are presented in Fig.~\ref{Experiment_InSb}. A well-marked suppression
of the specular reflectivity is observed in a close vicinity of the Rayleigh point, $\theta_R \simeq 19$ deg,
$\sin\theta_R = \lambda/d - 1$, where the $-1^{\text{st}}$ diffraction order transforms from the propagating wave
($\theta > \theta_R$) to the inhomogeneous one ($\theta < \theta_R$). All the reflectivity minima correspond to the
inhomogeneous $-1^{\text{st}}$ diffraction order that proves resonance SPP excitation. Note,
that the reflectivity minima for the gratings with the residual photoresist are grouped near $\theta \approx 18$ deg,
while for the fourth (pure) grating, with the same profile and period as the third one has, the reflectivity
lies aside, at $\theta \approx 18.5$ deg. It is clear that this difference is due to presence of the photoresist
on the grating surface. Besides the resonance shift, the photoresist layer increases the resonance width. The resonance width
also increases with the grating depth increase.

A simple explanation of the experimental results follows from the analytical theory
\cite{Kats_2002, Kats_2007, Balakhonova_2007, Spevak_2011}. Let a plane $p$-polarized wave ,
$\overrightarrow{H} = \overrightarrow{e}_y\exp[ik(\alpha_0 x + \beta_0 z)]$, (with $\alpha_0 = \sin\theta$,
$\beta_0 = \cos\theta$, $k = \sqrt{\varepsilon_1}\omega/c$, where $\varepsilon_1$ is the dielectric permittivity of the
upper media, see Fig.~\ref{Geometry}), be incident onto a periodically profiled surface, $z = \zeta(x)$,
$\zeta(x) = \sum_n \zeta_n \exp(ingx)$, with $g = 2\pi/d$, where $d$ is the grating period. Then, the amplitudes $h_n$ of the
diffracted waves, $h_n \exp[ik(\alpha_n x - \beta_n z)]$, where $\alpha_n = \alpha_0 + ng/k$,
$\beta_n = \sqrt{1 - {\alpha_n}^2}$ with $\Re, \Im(\beta_n) \geq 0$, are
\begin{equation}\label{1}
    h_r = (1 + R)\nu_{r0}/\Delta_r,    \quad    \Delta_r = \beta_r + \xi + \Gamma_r,
\end{equation}
\begin{equation}\label{2}
    h_N = \delta_{N0}R + (\nu_{N0} + \nu_{Nr}h_r)/(\beta_N + \xi),
\end{equation}
where $R = (\beta_0 - \xi)/(\beta_0 + \xi)$ is the Fresnel reflection coefficient from a plane interface,
$\nu_{mn} = ik(1 - \alpha_m \alpha_n)\zeta_{m-n}$ are proportional to the Fourier harmonics of the grating,
$\Gamma_r = \sum_{N \neq r}{|\nu_{Nr}|^2/(\beta_N + \xi)}$.
Indexes $r$ and $N$ correspond to the resonance (in our case $r = -1$) and nonresonance diffraction orders,
respectively, $\xi = \sqrt{\varepsilon_1/\varepsilon}$  is the relative surface impedance,
$\varepsilon$ is the dielectric permittivity of the semiconducting medium. The quantity $\Gamma_r$ in the
denominator of Eq.~(\ref{1}) takes into account scattering of the resonance diffraction order, $h_r$,
into various nonresonance orders, $h_N$, and inversely, the scattering of the latter back to  $h_r$,
\begin{equation}\label{3a}
  h_r
    \xrightarrow[]{\zeta_{N-r}}  \\
     h_N   \\
     \xrightarrow[]{\zeta_{r-N}}
    h_r .
\end{equation}
It is supposed that $\Gamma_r$ is a finite quantity owing to the fast decrease of the profile Fourier harmonics, $\zeta_n$,
with $|n|$ increase. Generally speaking, the quantity $\Gamma_r$ depends on the incidence angle $\theta$.
But this dependence is weak, the relative variation is no more than 5\% in the limits of the resonance width even
for the deepest grating. Therefore, in the first approximation we can take $\Gamma_r$ at the Rayleigh angle,
$\theta = \theta_R \approx 19$ deg.
\begin{figure}[!t]
    \centerline{\scalebox{0.45}[0.45]{\includegraphics{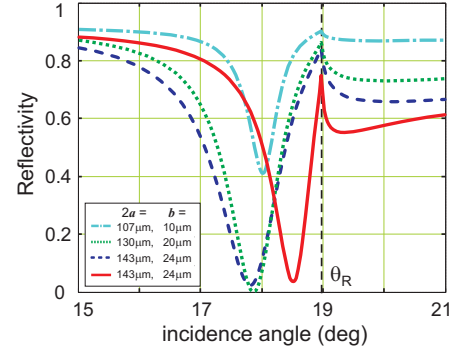}}}\caption[]
    {Dependence of the specular reflectivity on the incidence angle: theoretical results for a plane wave.}
    \label{Theory_InSb}
\end{figure}

The resonance condition, resulting in maximum of $|h_r|^2$ and corresponding minimum of $|h_0|^2$, is
$\Im \Delta_r = 0$, or explicitly
\begin{equation}\label{3}
    \theta_{\text{res}} = \arcsin\left\{ - \frac{r\lambda}{\sqrt{\varepsilon_1}d} + \frac{r}{|r|}\sqrt{1 + \Im(\xi + \Gamma_r)^2}\right\}.
\end{equation}
Emphasize, Eq.~(\ref{3}) derived from the theory for a plane wave
gives the resonance position which is in a good agreement with the experimental one for the pure grating,
cf. red curves in Fig.~\ref{Experiment_InSb} and Fig.~\ref{Theory_InSb}.

The full resonance width at the half-height can be obtained from Eq.~(\ref{1}) and is
\begin{equation}\label{3a}
    \Delta\theta = \frac{2 \Re(\xi + \Gamma_r) |\Im(\xi + \Gamma_r)|}{\cos\theta}.
\end{equation}
According to Eq.~(\ref{3a}), $\Delta\theta = 0.7$ deg for the pure grating, while the experimental value is
$\Delta\theta = 1.15$ deg. Evidently, the greater resonance width in the experiment is
conditioned by the angular divergence of the incident laser beam, $\delta\theta \approx 0.9$ deg.

Note here that the dispersion relation of the SPP on a plane surface of a highly conducting medium,
$| \varepsilon | \gg 1$, is $q^2 = k^2 (1 - \xi^2)$, $\xi = \xi(\omega)$, where $q$ is the SPP wavenumber.
It can be strictly obtained from the above general solution, Eq.~(\ref{1}), together with the resonance condition,
Eq.~(\ref{3}), represented in the form $k\sin\theta + rg = \textrm{sign}(r) \Re(q)$, that the periodic
corrugations result in renormalization of the surface impedance,
\begin{equation}\label{3b}
   \xi \rightarrow \overline{\xi} = \xi + \Gamma_r.
\end{equation}
This renormalization leads to the shift and broadening of the SPP resonance
as compared to those for a plane surface: the deeper the groves the greater the resonance shift and width.
It is just the renormalized impedance,
$\overline{\xi} = \xi + \Gamma_r$, stands in the denominator in Eq.~(\ref{1}).
It should be noted that $\Re\Gamma_r > 0, \Im\Gamma_r < 0$, that is their signs coincide with those for the surface
impedance, $\Re\xi > 0, \Im\xi < 0$.

\begin{figure}[!t]
    \centerline{\scalebox{0.4}[0.4]{\includegraphics{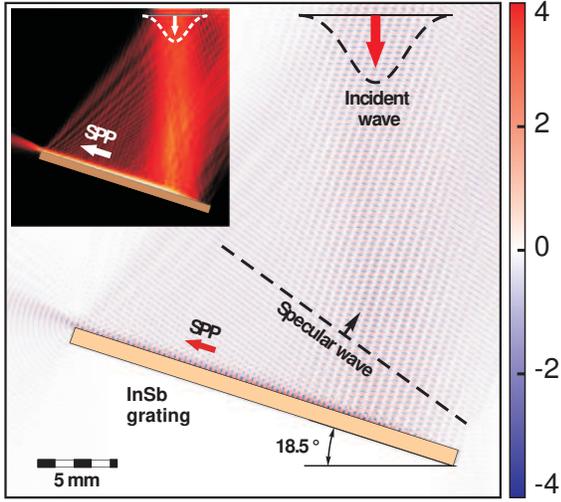}}}\caption[]
    {The space distribution of the magnetic field at resonance diffraction
    of the restricted beam on the pure grating.
    The angle of incidence  $\theta = 18.5^{\circ}$
    corresponds to the specular reflectivity minimum.
    The main picture presents the instantaneous field distribution,
    and the time averaged intensity distribution is shown in the inset.}
    \label{Space}
\end{figure}

To explain the additional shift and broadening of the SPP resonance due to the residual photoresist, we assume
that the photoresist covers the grating by a thin continuous layer. Supposing that the effects of the covering thin layer
and corrugations are independent, we can use the dispersion relation for the SPP on a
plane surface with a thin layer \cite{Raether_1988}. Corresponding consideration shows that the effect of the thin layer
of the thickness $\ell$ with the dielectric permittivity $\varepsilon_2$ is equivalent to the effective impedance
\begin{equation}\label{3c}
   \xi \rightarrow \tilde{\xi} = \xi + G, \ \ \
    G = -i (1 - \frac{\varepsilon_2}{\varepsilon})
    (1 - \frac{\varepsilon_1}{\varepsilon_2})k \ell,
\end{equation}
provided $|\varepsilon| \gg \varepsilon_1$, and
$\sqrt{|\varepsilon_2/\varepsilon_1 - 1|} \ k \ell \ll 1$.
There are no other limitations on $\varepsilon_2$ value, and the ratio ${\varepsilon_2}/{\varepsilon}$ can be arbitrary.
Note, the factor $G$ vanishes, $G = 0$, if $\varepsilon_2 = \varepsilon_1$ or $\varepsilon_2 = \varepsilon$.
For $\varepsilon_2 > \varepsilon_1$ the dielectric layer produces the shift and broadening in the same direction
as the corrugations do since $\Re G > 0 $, $\Im G < 0$.

In the case of the corrugated surface covered with a layer the plasmon-polariton dispersion relation
can be obtained by taking into account the both independent factors simultaneously,
$\xi \rightarrow \xi + \Gamma_r + G$. Correspondingly,
in the diffraction problem under examination we have to change the resonance denominator as follows,
\begin{equation}\label{6}
   \Delta_r \rightarrow \beta_r + \xi + \Gamma_r + G.
\end{equation}
The results received in this way are presented
in Fig.~\ref{Theory_InSb} for $\varepsilon_1 = 1$, $\varepsilon_2 = 2.6$ and the layer thickness $\ell = 6~\mu \text{m}$
(this value gives the best agreement with the experimental data).

An analogous transformation must be performed in Eqs.~(\ref{3}), (\ref{3a}) resulting in
\begin{equation}\label{7}
    \theta_{\text{res}} = \arcsin\left\{ - \frac{r\lambda}{\sqrt{\varepsilon_1}d} + \frac{r}{|r|}\sqrt{1 + \Im(\xi + \Gamma_r + G)^2}\right\},
\end{equation}
\begin{equation}\label{8}
    \Delta\theta = \frac{2 \Re(\xi + \Gamma_r + G) |\Im(\xi + \Gamma_r + G)|}{\cos\theta}.
\end{equation}
To make the picture more complete we present here the characteristic values:
$\xi = 0.021 - 0.101i$, $G = 0.0008 - 0.071i$, and $\Gamma_r(\theta_R) = 0.004 - 0.007i; 0.021 - 0.019i; 0.031 - 0.024i$
for the three successive gratings. The imaginary parts of these quantities determine the resonance position, see Eq.~(\ref{7}).
The resonance shift caused by the thin covering layer amounts to $0.6-0.7$ deg depending on the groove depth,
and it prevails the shift produced by corrugations.
In turn, the real parts of the $\xi$, $G$, $\Gamma_r$ are responsible for the losses (the active or radiative one)
and, consequently, determine the resonance width, see Eq.~(\ref{8}). Although the
layer is transparent ($\varepsilon_2$ is real) and does not absorb the radiation itself,  the field tightening
to the grating surface due to the layer results in additional grating absorption and considerable broadening of the resonance.
Analytically it is manifested by the product of the real and imaginary parts of the renormalized impedance, see Eq.~(\ref{8}).
Thus, for the deepest grating with the photoresist layer
the resonance width is about $1.0$ deg, as opposed to $0.7$ deg for the pure grating; meanwhile, in the experiment these
parameters are $1.5$ deg, and $1.15$ deg, respectively, because of the laser beam divergence.

As one can see, the theory with the renormalized resonance denominator, Eq.~(\ref{6}), successfully describes the resonance
shift conditioned by the covered layer and corrugations. However, the resonance width  for a plane wave is evidently smaller
than the experimental one for the restricted beam. To estimate the latter theoretically we have to take into account the angular width
of the incident beam. Such treatment (not presented here) gives satisfactory agreement with experimental data for the
resonance width as well.

To illustrate qualitatively the resonance diffraction of the restricted laser beam the space dependence of the magnetic
field, calculated with COMSOL, is presented in Fig.~\ref{Space}. The picture presented corresponds to the
$p$-polarized plane Gaussian beam with the same divergence as in the experiment impinging on the pure semiconductor grating
at the resonance angle. Both the instantaneous field and time averaged intensity patterns are presented in the picture.
Emphasize, the well pronounced field amplification near the grating surface confirms the SPP excitation.
Within the laser spot area the SPP magnitude increases in the propagation direction (to the left).
It can be also seen that SPP runs outside the laser spot limits  -- according to the estimation the free
path length constitutes $\sim 25 \lambda \approx 8.4$~mm. Outside the spot the propagating SPP is rescattered
in the specular direction. One can also notice a negative displacement of the reflected beam as a whole,
a so called Goos-H\"{a}nchen effect \cite{Bonnet_2001, Bliokh_2013}.

In conclusion, we have investigated both theoretically and experimentally resonance suppression of the specular reflection
of the THz laser radiation from semiconductor gratings. It was shown that covering the grating with a thin transparent layer
results in the SPP resonance shift and broadening. The theoretical relations obtained can be used for studying
the surface properties and agents deposited on the grating surface.

This work was supported by the Ukrainian State program "Nanotechnologies and nanomaterials", and
by the program of the National Academy of Sciences of Ukraine "Fundamental problems of nanostructured systems,
nanomaterials and nanotechnologies".

\end{document}